\documentclass{article} 
\usepackage{tech_report,times}
\usepackage{graphicx}
\usepackage{amsmath,amssymb} 
\usepackage{color}
\usepackage{todonotes}
\usepackage{hyperref}
\usepackage{multirow}

\title{Connecting Compression Spaces with Transformer for Approximate Nearest Neighbor Search}

\begin{document}

\maketitle

\begin{abstract}
We propose a generic feature compression method for Approximate Nearest Neighbor Search (ANNS) problems, which speeds up existing ANNS methods in a plug-and-play manner. Specifically, based on transformer, we propose a new network structure to compress the feature into a low dimensional space, and an inhomogeneous neighborhood relationship preserving (INRP) loss that aims to maintain high search accuracy. Specifically, we use multiple compression projections to cast the feature into many low dimensional spaces, and then use transformer to globally optimize these projections such that the features are well compressed following the guidance from our loss function. The loss function is designed to assign high weights on point pairs that are close in original feature space, and keep their distances in projected space. Keeping these distances helps maintain the eventual top-k retrieval accuracy, and down weighting others creates room for feature compression. In experiments, we run our compression method on public datasets, and use the compressed features in graph based, product quantization and scalar quantization based ANNS solutions. Experimental results show that our compression method can significantly improve the efficiency of these methods while preserves or even improves search accuracy, suggesting its broad potential impact on real world applications.
\end{abstract}

\section{Introduction}

Approximate nearest neighbor search (ANNS) methods focus on searching for $k$ approximate nearest neighbors from a given database to a given query node $q$. It is a fundamental technology in information retrieval and is widely used in applications such as search engines and recommendation systems. The common goal of ANNS approaches is to minimize the search latency while maintain a low search accuracy loss on a fixed hardware constraint. 

As stated in~\cite{subramanya2019diskann},  currently, the two most popular ANNS methods are graph based approaches and quantization based approaches. Many of these methods, such as product quantization (PQ) \cite{jegou2010product}, HNSW \cite{malkov2018efficient} and NSG \cite{fu2017fast} are widely used in real-world applications. Although popular, there are rooms to further improve them. In PQ, the sub vectors are quantified into codewords using a clustering objective, which is a poor proxy to search accuracy. In graph based methods, there are a large number of distance computations in index building time (for instance, indexing complexity NSG graph is ${O}(kn^{\frac{1+d}{d}}\log{}n^{\frac{1}{d}}+n\log{}n)$~\cite{fu2017fast}, where $d$ is data dimension), which in real world applications can last weeks on billion scale datasets. For PQ related methods, introducing the feature compression as an intermediate step avoids direct feature quantization from high dimensional space. Such a two stage quantization strategy usually results in higher accuracy. For graph based methods, computing distance in low dimensional space reduces computational cost linearly, which significantly speeds up indexing.

Existing feature compression methods such as principal components analysis (PCA) \cite{wold1987principal}, Variational Auto-Encoders(VAE)~\cite{pu2016variational} focus on keeping the information in the input features instead of their neighborhood relations, but in ANNS applications the neighborhood structure affects the eventual accuracy more than the locations of the individual data points themselves. As is shown in experiments, directly applying these methods lead to significantly reduced search accuracy, which is not suitable for applications requiring both high speed and high recall.

In this paper, we propose a feature compression method for the approximate nearest neighbor search problem, which aims to retain the local neighborhood relations instead of the fidelity of the reconstructed features. Our method is composed of a compression network which connect compression spaces with transformer (CCST), and an inhomogeneous neighborhood relationship preserving (INRP) loss. 

Our CCST is a combination of projection units that projects original features to low dimensions, and transformer units that compose these projections to generate the output feature. The projection unit is initialized as sparse random projection (SRP) \cite{li2006very}, which is proved by JL (Johnson–Lindenstrauss) lemma that this projection reserves distances of an increasing number of data points with decreasing dimension reduction ratios. We design our network with multiple projections units, so that each can be used to compress a local neighborhood, and the entire space can be covered in a piece-wise manner by the ensemble of these units. We then use transformers to adaptively combine the output of these random projection units, because the transformer unit has the well known property of globally attending its inputs, which could lead to better global alignment of these projections. In other words, multiple compression spaces are connected to one via transformer. Different from standard transformer based networks, we design an input dependent, non-trivial output token named compression token, which is itself a compressed feature derived from the input feature, and is updated by the transformer layers with skip connections. The compression token is designed to provide an anchor vector to the multi-head attention units so that transformers properly mix the output of random projection units to generate the output compressed feature.

Our training loss is designed for the purpose of preserving local neighborhood structures. In most ANNS applications, user is mostly interested in the accuracy of the top few nearest neighbors. So, changes in the distances between a query and its far away points will not affect the search result, as long as their distances are still large. Inspired by this property in ANNS problems, we design the INRP loss, which assigns high losses on point pairs whose distances are within a threshold. This loss design allows certain information of the original feature to be discarded while still maintaining high retrieval accuracy.

Our experiments show that our feature compression model can be seamlessly used with popular ANNS methods. While saving $1/2$ to $3/4$ indexing time, graphs built using our compressed features can improve the recall slightly for HNSW and NSG methods. It also significantly improves recalls of 1@1 and recalls of 1@5 metrics by more than 10.0 percentage points for PQ based methods, and about 1.0 or 2.0 percentage points for a quantization methods tailored to HNSW methods ~\cite{douze2018link}.

Our main contributions are summarized as follows. 

\begin{enumerate}

\item We propose a novel feature vector compression model CCST for ANNS problem. In CCST, traditional projection based compression units and emerging transformer units are jointly used to compress features for ANNS problems. To our knowledge, this is the first attempt to apply transformer to this kind of task.

\item We propose an INRP loss that mainly keeps the distances of a point and its close local neighbors. This maintains top $k$ retrieval accuracy and creates space for lossy feature compression.

\item The experiment results show that our proposed method can be used in most ANNS methods to improve efficiency. It speeds up indexing speed to 2$\times$ to 4$\times$ for graph based methods and it improves recalls significantly for PQ related methods.

\end{enumerate}

\section{Related Work}

\subsection{ANNS methods}

Existing ANNS approaches can be divided into four main types: 1) tree-structure based strategies \cite{silpa2008optimised}, which partitions indexed datapoints into different subspaces based on specific conditions; 2) Locality sensitive hashing (LSH) related methods, which map similar items to the same symbol with a high probability \cite{andoni2015optimal}; 3) product quantization (PQ) related approaches decompose the space into a Cartesian product of low dimensional subspaces and quantize each subspace separately \cite{jegou2010product}; 4) graph based frameworks search on pre-built relative neighborhood graphs (RNG) to find the closest data points to query \cite{malkov2018efficient}. Compared with PQ and graph based methods, tree based and LSH based methods need ensembles of trees or hash tables to achieve similar accuracy, which consume considerably more memory and are less frequently used in large scale ANNS problems.

Product quantization based approach reduces the index memory cost by holding quantized codes and speeds up distance computation using online or offline computed distance look up tables. When combined with inverted index to further reduce the number of distance computations, PQ becomes a strong baseline for modern ANNS systems. Compared with quantization based approach, which partitions original feature space to generate shorter features, our method directly maps the original features into a new space. These two method are orthogonal. As is shown in experiments, the new space can be further quantized by PQ, which generates even better speed and accuracy trade-offs than using PQ method alone.

Graph based algorithms usually construct a navigable graph over database, and searches along the edges of this built graph to find the closest points to query. The most recent works in this category are NSG \cite{fu2017fast}, HNSW \cite{malkov2018efficient}, Disk-ANN \cite{subramanya2019diskann} and HM-ANN \cite{ren2020hm}. NSG builds a relatively sparse indexing graph to reduce memory usage and improve search speed. HNSW employs hierarchical graph structure to reduce query latency. Disk-ANN and HM-ANN focus on reducing memory overhead via adopting quantization and heterogeneous memory. Besides, Douze et al. proposed Link and Code (L$\&$C), which takes HNSW as basic ANNS framework and replaces full precision vectors with refined quantiztion codes \cite{douze2018link}. Compared with brute force, these graph based methods visit 1/1000 or even less of the indexed points in each search, significantly reduce the number of needed distance computations. However, most of these methods suffer from the problem of high indexing time, as constructing a navigable graph over database has high time complexity. For instance, ${O}(n\log{}n)$ for HNSW and ${O}(kn^{\frac{1+d}{d}}\log{}n^{\frac{1}{d}}+n\log{}n)$ for NSG. In real world application, building a navigable graph over a billion-scale database with 30+ threads costs several days or even weeks. 

\subsection{Dimension reduction / compression}

In early stage, almost all dimension reduction methods are based on mathematical theory. PCA \cite{wold1987principal} and independent components analysis (ICA) \cite{hyvarinen2000independent} are proposed to transform vectors into lower dimensional space, where the most information of high dimension vectors are retained. Random projection~\cite{achlioptas2003database}\cite{li2006very} project data into low dimensional space, while approximately preserving structure information in original feature space. The efficiency of random projection methods is guaranteed by Johnson-Lindenstrauss (JL) lemma \cite{johnson1984extensions}. This lemma states that data points in sufficiently high  dimensional space may be projected into suitable low dimensional space while approximately preserving the distances between the points in original space. 

In 2006, Hinton and Salakhutdiov proposed to compress high dimensional data via neural network \cite{hinton2006reducing}, and several recent works on dimension reduction have shifted their methods to neural networks. For example, VAE maps data to low dimensional space and forces the encoded features to follow a multivariate Gaussian distribution \cite{pu2016variational}. 

Very recently, Sablayrolles et al. \cite{sablayrolles2019spreading} proposed an end-to-end quantization model for similarity search, where a compression network named catalyst is designed and trained to compress vectors from ${d}_{in}$-dimensional input space to the hypersphere of a ${d}_{out}$-dimensional space, where ${d}_{in}>{d}_{out}$. Our proposed CCST is close to this work, as it involves an ANNS oriented loss and a learnable network to compress feature vectors. Compared with catalyst, our training objective is easier to implement as it does not involve offline exhaustive search for positive and negative pairs. Also, we design a deep compression model for the ANNS problem, where \cite{sablayrolles2019spreading} uses multi-layer perceptions to demonstrate their entropy maximization idea.

\subsection{Transformer}
Vaswani et al. first proposed transformer model to solve NLP problem in \cite{vaswani2017attention}. Inspired by the promising performance of transformer, several works started to introduce transformer into vision tasks and proposed a new type of models. Dosovitskiy et.al proposed ViT, in which the input image is cropped into a sequence of patches to meet the input format requirement of transformer \cite{dosovitskiy2020image}. Later, Touvron et al. proposed to use knowledge distillation to overcome the difficulties of training ViT models  \cite{touvron2021training}. Liu et al proposed a hierarchical transformer, where representation is computed within shifted windows \cite{liu2021swin} to reduce computation cost. Very recently, Graham et al. mixed CNN and transformer in their LeVit model, which significantly outperforms previous CNNs and ViT models with respect to the speed/accuracy tradeoff \cite{graham2021levit}. 

The transformer part of our CCST model is inspired by the works above. We have made several modifications to better fit transformer with our applications, which are detailed in section \textit{global optimization part}.

\begin{figure}[t]
\begin{center}
\includegraphics[width=4.6 in]{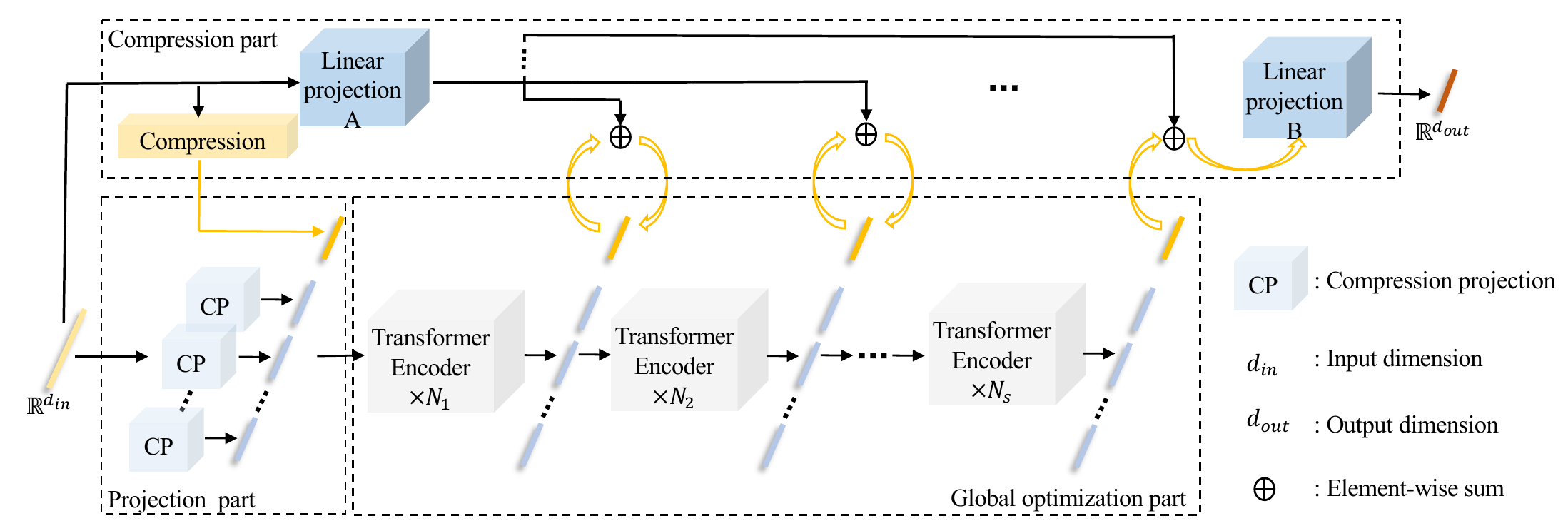}
\end{center}
\caption{Overview of the CCST model. The proposed model consists of three major parts, including compression part, projection part and global optimization part. The projection part projects input vectors into multiple subspaces and outputs a sequence of sub vectors. Global optimization part optimizes the sequence of sub vectors from global perspective and summarizes all optimized information into compression token. Compression part provides initial the compression token, and interacts with global optimization part to generate the final compression result.}
\label{fig:framework}
\end{figure}

\section{The Proposed Approach}

In this section, we present our proposed method in detail. We first introduce the architecture of the proposed CCST. Then we elaborate on our INRP loss function. 

\subsection{CCST}
As illustrated in Fig.~\ref{fig:framework}, the proposed CCST consists of three major parts, including projection part, global optimization part and compression part. 

The projection part casts the input feature vector $x\in \mathbb{R}^{d_{in}}$ to $n$ different low dimensional spaces. The output of this part is a sequence of low dimensional vectors $\left[{p}^{1}(x), {p}^{2}(x), \cdots {p}^{n}(x)\right]$, where ${p}^{i}(x)\in \mathbb{R}^{d_{out}}, i=1, \cdots, n$. $d_{in}$ and $d_{out}$ denote feature dimensions before and after compression, respectively. $p^{i}()$ is a compression projection function. $p^{i}(x)=W^{i}x$, where $W^{i} \in \mathbb{R}^{d_{in}\times d_{out}}$. The global optimization part takes $\left[cp(x), {p}^{1}(x), {p}^{2}(x), \cdots {p}^{n}(x)\right]$ as input and globally optimizes all sub vectors step by step. The global optimization part then summarizes all optimized sub vectors into a compression token $cp(x)$. The compression part is responsible for initializing the compression token, interacting with global optimization part in each step and further processes the information in compression token to generate the final compression result $f(x)\in \mathbb{R}^{d_{out}}$. 

\subsubsection{Projection part}

Our goal is compressing feature vectors into a low dimensional space, where the data neighborhood relation in original space is preserved. It seems that random projections satisfy our requirement and there are already some classical random projection methods. Unfortunately, according to JL lemma~\cite{johnson1984extensions}, the projection error of using single random projection function satisfies:
\begin{equation}
(1-\epsilon){||{x}_{i}-{x}_{j}||}^{2}_{2} \leq {||p({x}_{i})-p({x}_{j})||}^{2}_{2} \leq (1+\epsilon){||{x}_{i}-{x}_{j}||}^{2}_{2}
\label{ieq:jl-lemma}
\end{equation}
where $x_{i}$ and $x_{j}$ represent two data points in high dimensional space and $p(x_{i})$ and $p(x_{j})$ are corresponding projection data points in low dimensional space. $p()$ is the projection function. $\epsilon$ satisfies equation $d_{out}>\frac{4\ln(m)}{\epsilon^{2}/2 - \epsilon^{3}/3}$, where $0<\epsilon <1$ and $m$ represents data size. Random projections are inapplicable in our case. For example, if compressing 960 dimensional vectors in GIST1M dataset to 480 dimensions, we have $0.63<\epsilon<1$. In GIST1M, the distance of a query to its nearest neighbor and to its hundredth neighbor are probably 1.177 and 1.5615, respectively. After projection, with the minimum $\epsilon$, the distance of a query to its nearest neighbor and to its hundredth neighbor fall into the scope of [0.7160, 1.5027] and the scope of [0.9498, 1.9936], respectively. These two scopes have a considerable overlap, which may disturb neighborhood relationship and lead to a lower search accuracy. We can draw a consistent conclusion from our experimental results. 


Existing random projection methods do not address our problem directly, but they give us some inspirations. Here, we propose to use $n$ different projections to boost compression accuracy. Specifically, we initialize the $n$ projection matrices in ${p^i(x)=W_i x}$ ($i=1, 2, \cdots, n$) on input $x\in \mathbb{R}^{d_{in}}$ to generate a sequence of features $\left[{p}^{1}(x), {p}^{2}(x), \cdots {p}^{n}(x)\right]$. Following~\cite{li2006very}, the elements of in each $W_i$ are randomly drawn from:

\begin{equation}
\left\{
             \begin{array}{crr}
             -\sqrt{\frac{s}{d_{out}}} & \text{with probability} & 1/2s, \\
             0 & \text{with probability} & 1-1/s \\
             \sqrt{\frac{s}{d_{out}}} & \text{with probability} & 1/2s \\
            
             \end{array}
\right.    
\end{equation}
where $s=\sqrt{d_in}$. Different from traditional random projections, projection matrices used in here will be further optimized by our INRP loss. 

\begin{figure}[t]
\begin{center}
\includegraphics[width=3 in]{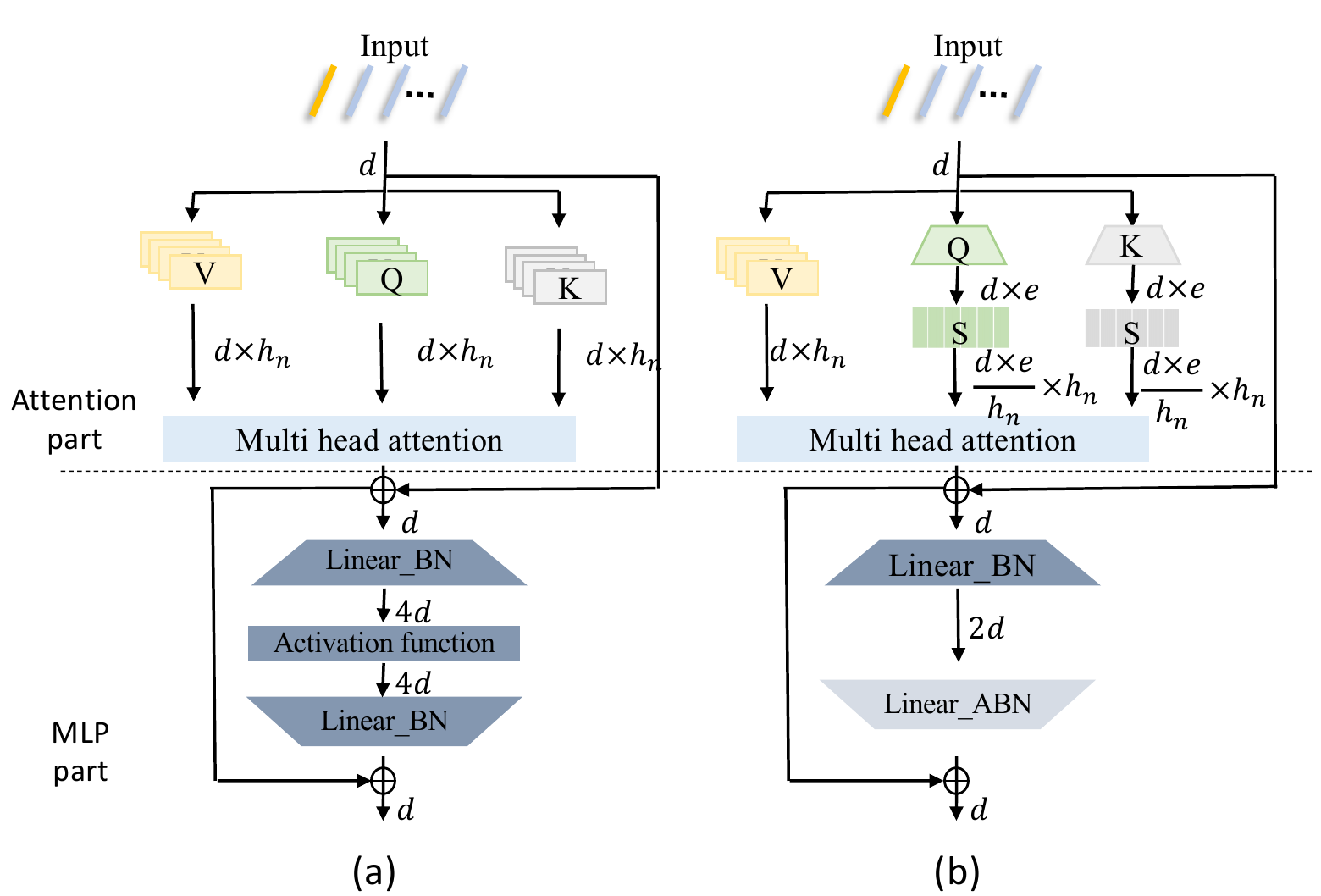}
\end{center}
\caption{Different modules. (a) Transformer module of ViT~\cite{dosovitskiy2020image}. (b) Transformer module proposed in this paper. By adopting lightweight designs, transformer module in (b) has fewer parameters than that in (a). From (a) to (b), the number of parameters in attention part decreases from $4d^{2}h_{n}$ to $2d^{2}h_{n}+2d^{2}e$ and the number of parameters in MLP part reduces from $8d^{2}$ to $4d^{2}$, where expansion factor $e$ is smaller than the number of heads $h_{n}$.}
\label{fig:transformer}
\end{figure}

\subsubsection{Global optimization part}
The projection part outputs a sequence of low dimensional vectors, which brings in the core problem of this global optimization part, that is how to generate the result compressed feature using these features from different sub spaces. 

Here, as the problem that we need to solve is exactly what transformer model is good at, we tailor a transformer model to overcome the core problem. One biggest advantage of transformer is it good at capturing and taking use of relationships between different tokens to get a global optimization result. When we treat features from different sub spaces as different tokens, our core problem becomes optimizing a sequence of tokens from the global perspective. So, based on ViT \cite{dosovitskiy2020image} structure, we propose a new transformer model for feature compression. 

Based on the characteristics of our compression problem, we have made four major modifications to the original ViT structure, including two structural modifications and two lightweight designs.

\textit{Structural modifications} ensure the new proposed transformer structure is aligned with our problem. These two modifications are:
\begin{enumerate}
\item Discard position embedding. In transformer models for processing sentences or images, position embedding and relative position embedding are widely used, as the order of words and the positions or relative positions of images patches contain important information. However, in our case, the order of random projections is fixed. Naturally, position embedding is discarded in our transformer model. Interestingly, when we add position embedding into our transformer, the search accuracy drops. We conjecture this is because randomly initialized position coding disturbs our model.   
\item Add compression token. In our transformer model, an extra token named compression token is used. Structurally, this design is similar to the transformer model in ViT, where an extra token named classification token is also employed. Compression token and classification token are added for different purposes and they have different functions. Compression token provides a base reference for optimizing and summarizing information from different projection spaces and it also works as a bridge between compression part and global optimization part. Classification token is used as a placeholder to perform classification. Different from classification token, which is a learned constant, our compression token is derived from input feature vector. Initializing compression token with random numbers is not in accordance with its role of working as base reference. 
\end{enumerate}

\textit{Lightweight designs} are adopted to reduce the number of learnable parameters. As shown in Fig.~\ref{fig:transformer}, two lightweight designs for MLP part and attention part are: 

\begin{enumerate}
\item In MLP part, we decrease the expansion ratio (changes among widths of the adjacent linear mapping layers) from 4 to 2 to reduce the number of parameters by half, as shown in the bottom half of Fig.~\ref{fig:transformer} (b). In addition, the basic computing unit in our model is Linear\_ABN, which, in data processing order, consists of a linear mapping layer, activation function and batch normalization. As pointed out in~\cite{chen2019rethinking}, placing batch normalization layer before ReLU leads to the conv layer updated in a suboptimal way due to the nonnegative responses of activation function. So we design our unit following the order of conv$\rightarrow$activation$\rightarrow$bn. This order is also recommended in~\cite{zhuang2019structured}. 

\item Attention part has three intermediate variables, which are represented as Value ($V$), Query ($Q$) and Key ($K$). The output of head $i$ is calculated as:
\begin{equation}
\mathrm{head}_{i}(Q, K, V)= \mathrm{softmax}(\frac{QW_{i}^{Q}\cdot {KW_{i}^{K}}^{T}}{\sqrt{d}})VW_{i}^{V}
\label{eq:attention}
\end{equation}
where $W_{i}^{Q}$, $W_{i}^{K}$ and $W_{i}^{V}$ are three projection matrices. As in Eqn.~\ref{eq:attention}, V stores main information, $Q$ and $K$ are used to generate attention weights. In ViT, $Q$, $K$ and $V$ are all the same in each head. Inspired by~\cite{vaswani2017attention}, we decrease the dimension of $Q$ and $K$ for reducing parameters of $W_{i}^{K}$ and $W_{i}^{V}$. 

\end{enumerate}

Structurally, as shown in the middle part of Fig.~\ref{fig:framework}, global optimization part consists of $s$ stages, each of which has $N_{i}$ transformer encoders. From stage $1$ to $s-1$, projected vector from compression part is added to compression token at the end of each stage. Finally, $cp(x)$ is from the output of last stage is fed back to compression part to generate the final compression result.

\subsubsection{Compression part}

The compression part is responsible for providing initial compression token, interacting with global optimization part in each step and generating the final compression result. Corresponding to three responsibilities three learnable modules are constructed in this part. As shown in the top part of Fig.~\ref{fig:framework}, these three modules are a compression module and two linear projection modules. 

The compression module is made up of a sinlge Linear\_ABN module. Each linear projection module has a linear mapping function. Compression module takes $x\in \mathbb{R}^{d_{in}}$ as input and outputs $cp(x)\in \mathbb{R}^{d_{out}}$. Linear projection $A$ maps input $x\in \mathbb{R}^{d_{in}}$ to $d_{out}$-dimensional vector, which is added to compression tokens of global optimization part. Linear projection $B$ takes compression token from the last stage of global optimization part as input and output the final compression result.

\subsection{INRP loss}

\begin{figure}[t]
\begin{center}
\includegraphics[width=3.4 in]{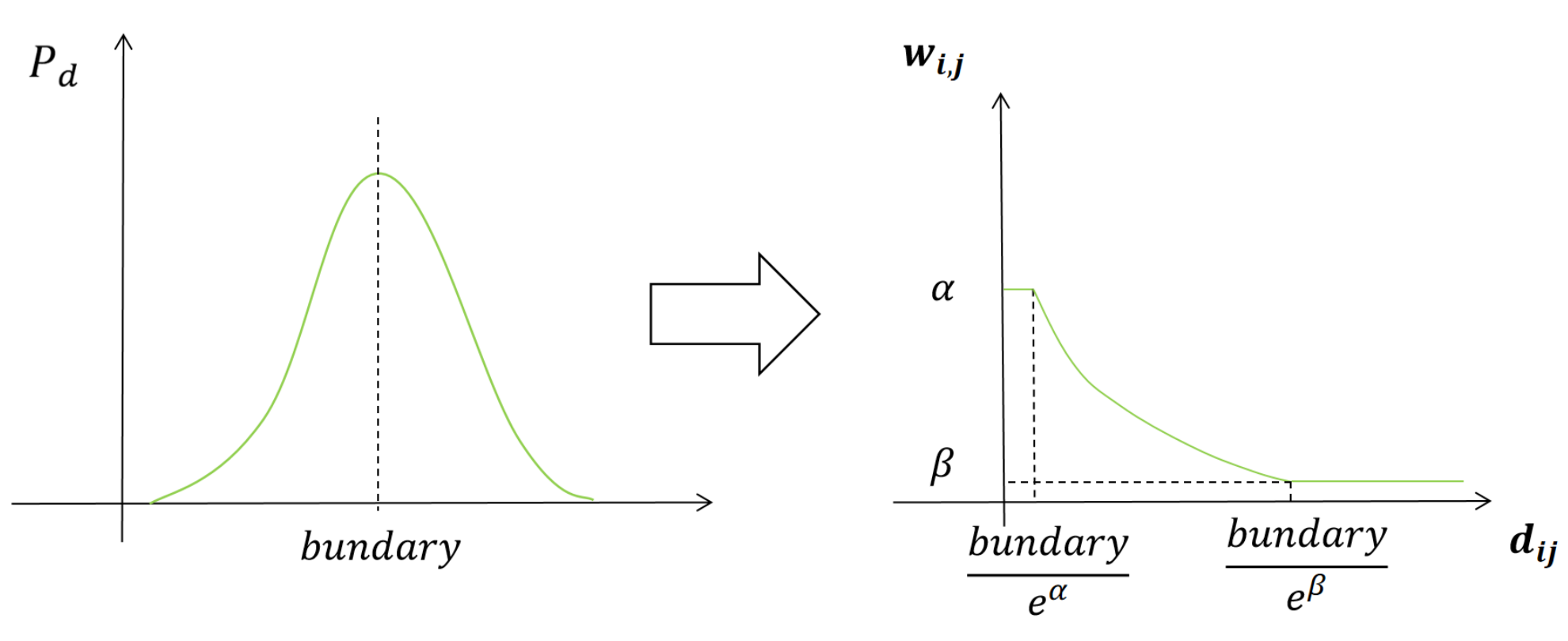}
\end{center}
\caption{The weight curve of INRP loss. $d_{ij}$ is Euclidean distance between $x_i$ and $x_j$.} 
\label{fig:loss}
\end{figure}

To preserve neighborhood structures, a straightforward way is training the network to keep distances between all possible point pairs. However, the goal of ANNS is searching for top $k$ approximate nearest neighbors to a query point, which generally are points that are close to query point in the original space. In other words, focusing on keeping distances between close pairs is more important, as these pairs affect search result the most. In \cite{guo2020accelerating}, Guo et al. presented a similar idea, which is proposed for quantization instead of compression. Inspired by this, we design our INRP loss, the formula of which is:
\begin{equation}\label{loss}
loss = \frac{1}{m^2}\sum_{i=1}^{m}{\sum_{j=1}^{m}}w_{ij}\cdot\lVert \lVert f(x_{i})-f(x_{j})\rVert_2 - \lVert x_{i}-x_{j}\rVert_2 \rVert_2
\end{equation}

where $x_{i}$ and $x_{j}$ are a pair of any two nodes from original space. $m$ denotes data size. $f(x_{i})$ and $f(x_{j})$ are compression results. ${w}_{ij}$ is weight for this pair and it is calculated as:
\begin{equation}
w_{ij}=\min(\alpha, \max(\beta, -\ln(\frac{d_{ij}}{boundary})))
\end{equation}
where $d_{ij}=||x_i-x_j||_{2}$. $\alpha$ and $\beta$ are two hyper-parameters, which are set to 2.0 and 0.01 respectively. $boundary$ denotes the average distance between any two nodes in original space. The curve is shown in Fig.~\ref{fig:loss}. In implementation, we use all pairs inside a mini-batch to approximate Eqn.\ref{loss}, which significantly simplifies computation.


\begin{table}[t]
\footnotesize
\renewcommand\arraystretch{1.2}
\setlength\tabcolsep{0.5pt}
\begin{center}
\caption{Experiments of speeding up indexing for graph based methods. C\_F represents the feature compression factor. IND and SS denote indexing time and search speed. q/s is query per second. Experiments are performed on a server with two Xeon Gold 5218 CPUs. We build indices with 32 threads and perform search with a single thread. All the speed numbers are averaged from 20 cold runs to rule out random factors. Comp\_t is time spent in compressing 1M feature vectors (database) with trained CSSTs. Here, a single RTX2080Ti GPU is used for accelerating compressing process and the GPU memory usage is restricted to about 1GB.}
\begin{tabular}{c|cc|ccc|cc|ccc|cc}
\hline
\multirow{4}*{C\_F} & \multicolumn{10}{c|}{Methods} &   \\
\cline{2-11}
          &  \multicolumn{5}{c|}{HNSW}  &  \multicolumn{5}{c|}{NSG} & Comp\_t &  GPU \\
\cline{2-11}
          & IND & SS  & \multicolumn{3}{c|}{Recall} &  IND  &  SS   & \multicolumn{3}{c|}{Recall} &  (s) & (G) \\
          & (s)  & (q/s) &  1@1  &  1@10 & 100@100     &  (s)  &  (q/s)  &  1@1  &  1@10 & 100@100           &   \\
\hline
\multicolumn{13}{c}{GIST1M} \\
\hline
          1  &  733 & 182  & 97.40 & 100.00 & 94.39  & 1118  & 179  & 97.40  & 100.00  & 93.97 &  -    &  -   \\ 
          2  & 454 & 184  & 97.70 & 100.00 & 94.55 &  778    & 180  & 98.00  & 99.90   & 94.16  & 21.43 & 1.19    \\
          4  & \bf{245} & 186  & \bf{98.00} & 100.00 & \bf{95.27} &  \bf{645}    & 182  & \bf{98.30} & 100.00  & \bf{94.85}  & 10.00 & 1.17    \\
\hline
\multicolumn{13}{c}{Deep1M} \\
\hline
          1  & 300  & 861   & 99.50 & 100.00 & 95.23 & 505 & 820  & 99.7  & 100.00 & 94.53     &  -    &  -   \\
          2  &   144  & 870   & 99.50 & 100.00 & 95.46 &  246  & 822  & 99.6  & 100.00 & 94.69 & 5.15  & 1.09 \\
          4  &   \bf{100}  & 874   & \bf{99.80} & 100.00 & \bf{95.60} &  \textbf{209}  & 842  & \bf{99.6}  & 100.00 & \bf{94.92} & 4.64  & 1.06 \\
\hline
\end{tabular}
\label{Table: indexing}
\end{center}
\end{table}

\begin{table}[t]
\footnotesize
\renewcommand\arraystretch{1.2}
\setlength\tabcolsep{2pt}
\begin{center}
\caption{Experiments of speeding up indexing for real world database and billion-scale database.}
\begin{tabular}{c|c|c|c|cc}
\hline
\multirow{2}*{Datasets}  &  \multirow{2}*{C\_factor} & indexing  & Recall & Compressing & GPU       \\
          &            &  time(h)  &  1@1   & time (h)    & usage(G)  \\

\hline
\multirow{2}*{Bigann-1B} &   1    & 106  &   92.6  &  -  & - \\
                         &   2    & 51   &   92.8  & 1.01  &  1.24  \\
\hline 

\hline
\end{tabular}
\label{Table:bigdata}
\end{center}
\end{table}

\section{Experiments}

\subsection{Datasets and implementation details}

{\bf Datasest} We carry out experiments on two million-scale benchmark datasets GIST1M and Deep1M \footnote{Both are downloaded from \url{https://www.cse.cuhk.edu.hk/systems/hash/gqr/datasets.html} }. GIST1M consists of 1 million of 960-dimensional handcrafted feature vectors, which are extracted from images with GIST descriptors~\cite{oliva2001modeling}. Deep1M contains 1 million of 256-dimensional deep feature vectors, which are extracted by using GoogleNet~\cite{krizhevsky2012imagenet}.

\noindent
{\bf Training setting} Following~\cite{muja2014scalable}\cite{ gong2012iterative}, we take database as training set. We train our model for 2400 epochs with the Adamw optimzier~\cite{loshchilov2017decoupled}, where the initial learning rate and batch size are set to 1e-4 and 1024 respectively. We use poly learning rate policy with power of 0.9 to adjust the learning rate for every epoch.

\noindent
{\bf Platforms} We implement our CSST using Pytorch\footnote{\url{https://pytorch.org/}}  and conduct ANNS experiments based on Faiss\footnote{\url{https://github.com/facebookresearch/faiss/releases/tag/v1.7.1} }.

\subsection{Speeding up indexing for graph-based methods}
In this section, we focus on using our proposed CSST to speed up indexing of graph-based methods. Two most popular graph-based methods HNSW~\cite{malkov2018efficient} and NSG~\cite{fu2017fast} are employed as baselines. We construct all HNSW indices with M=48, efConstruction=512, and perform search with efSearch=100 on GIST1M and efSearch=200 on Deep1M. Following~\cite{fu2017fast}\cite{subramanya2019diskann}, we build all NSG indices using R=60, L=70 and C=500 and initialize K-NN graphs with NN-descent. On GIST1M, we initialize K-NN graph using GK=400, L=400, iter=12, S=15 and R=100. On Deep1M, we initialize K-NN graph using GK=200, L=200, iter=10, S=12 and R=100.

We conduct three groups of experiments with different compression ratios. In all three groups, full-dimensional vectors are used to search nearest neighbors.  Full-dimensional vectors, feature vectors compressed with a factor of 2 and 4 are respectively used in indexing. 

Experiment results listed in Table~\ref{Table: indexing} show that using compressed features triples indexing speed for HNSW and doubles indexing speed for NSG. Taking experiments based on HNSW as an example, using compressed features reduce indexing time from 733 seconds to 454 seconds and 245 seconds for GIST1M dataset and decreases indexing time from 300 seconds to 144 and 100 seconds for Deep1M. Experiments based on NSG show similar trends. 



Interestingly, using compression feature vectors slightly improves search accuracy. For instance, on GIST1M, using $4\times$-compressed vectors improves recall of 1@1 to 98.00\% and improves recall of 100@100 to 95.60, 0.5 and 0.37 percentage points higher that that of baseline respectively. We conjecture this is due to that using compressed feature vectors to build index introduces some extra links and these links improves search accuracy, just like extra links selected by select neighbors heuristic improves the accuracy of HNSW~\cite{malkov2018efficient}. Further investigation experiments exceeds the scope of this paper and will be conducted in our future work. 

Note that the speedup is scalable to larger datasets. As shown in Table~\ref{Table:bigdata}, for Bigann-1B\footnote{https://dl.fbaipublicfiles.com/billion-scale-ann-benchmarks/bigann/base.1B.u8bin} (1 billion 128-d hand drafted features), using our features reduces indexing time from 106 hours to 51 hours.

\begin{table}[]
\footnotesize
\renewcommand\arraystretch{1.2}
\setlength\tabcolsep{1.6pt}
\begin{center}
\caption{Fusion experiments. Bytes denotes the size of quantized feacture vectors. Comp. and Quant. represent compression and quantization methods, respectively. Following default settings in their source code~\cite{douze2018link}, we conduct experiments on L\&C~\cite{douze2018link}, except that the M of coder is set to 8 (This gets highest accuracy.). For PQ experiments, we adopt IVFADC (nlist=8) to avoid exhaustive search.}
\begin{tabular}{c|c|c|cccc}
\hline
\multicolumn{7}{c}{GIST1M} \\
\hline
\multirow{2}*{Bytes} & \multirow{2}*{Comp.} &  \multirow{2}*{ANNS} & Speed  & \multicolumn{3}{c}{Recall} \\
&    &   & (q/s)  & 1@1  &  1@5  &  1@50      \\
\hline

 60  &  -       & PQ  &  115       & 23.0   &  45.2  & 79.7     \\
 60  &  CSST     & PQ  &  117       & 39.5(\textcolor{blue}{16.5$\uparrow$}) &  72.8(\textcolor{blue}{27.6$\uparrow$})  & 97.3(\textcolor{blue}{17.6$\uparrow$}) \\
\cline{2-7}
 60  &  -       & L\&C  & 1536   & 31.4  &  48.9 & 54.2 \\
 60  &  CSST     & L\&C  & 1527   & 30.7(\textcolor{red}{0.7$\downarrow$})  &  50.2(\textcolor{blue}{1.3$\uparrow$}) & 55.1(\textcolor{blue}{0.9$\uparrow$}) \\
\hline
 30 &  -       & PQ &  246    &  15.2  &  31.0  &  62.4  \\
 30 &  CSST     & PQ &  251    &  20.9(\textcolor{blue}{5.7$\uparrow$})  &  43.4(\textcolor{blue}{12.4$\uparrow$})  &  79.7(\textcolor{blue}{17.3$\uparrow$})  \\
\cline{2-7}
 30 &  -       & L\&C  & 2000  & 23.6 & 43.0 & 53.9 \\
 30 &  CSST     & L\&C  & 1992  & 23.9(\textcolor{blue}{0.3$\uparrow$}) & 43.9(\textcolor{blue}{0.9$\uparrow$}) & 54.3(\textcolor{blue}{0.4$\uparrow$}) \\
\hline
\multicolumn{7}{c}{Deep1M} \\
\hline
 32 &  -       & PQ   & 236     &  32.3  & 67.5  & 95.9  \\
 32 & CSST      & PQ   & 240     &  53.3(\textcolor{blue}{21.0$\uparrow$}) & 88.5(\textcolor{blue}{21.0$\uparrow$})  & 100.0(\textcolor{blue}{4.1$\uparrow$}) \\
\cline{2-7}
 32 &  -       & L\&C  & 2000 & 47.9 & 72.3 & 76.4  \\
 32 & CSST      & L\&C  & 2020 & 48.7(\textcolor{blue}{0.9$\uparrow$}) & 74.1(\textcolor{blue}{1.8$\uparrow$}) & 78.3(\textcolor{blue}{1.9$\uparrow$})  \\
 \hline
 16 &  -       & PQ    & 520  &  16.8  & 38.6  & 73.6  \\
 16 & Catalyst & PQ    & 545  &  19.8(\textcolor{blue}{3.0$\uparrow$})  & 47.9(\textcolor{blue}{9.3$\uparrow$})  & 82.5(\textcolor{blue}{8.9$\uparrow$})  \\
 16 & CSST      & PQ    & 537  &  32.5(\textcolor{blue}{15.7$\uparrow$})  & 65.6(\textcolor{blue}{27.0$\uparrow$})  & 94.4(\textcolor{blue}{20.8$\uparrow$})  \\
\cline{2-7}
 16 &  -       & L\&C  & 2677 & 28.8 & 55.6 & 71.6 \\
 16 & CSST      & L\&C  & 2676 & 29.2(\textcolor{blue}{0.4$\uparrow$}) & 56.9(\textcolor{blue}{1.3$\uparrow$}) & 73.7(\textcolor{blue}{2.1$\uparrow$}) \\

\hline
\end{tabular}
\end{center}
\label{Table:fusion}
\end{table}

\subsection{Improving accuracy and speed for PQ related methods}
Besides graph-based methods, PQ related methods are also very popular in real world applications. As quantization and dimension compression are orthogonal at the method level, we conjecture that our proposed CSST can be applied on PQ related methods to achieve higher accuracy. In this section, we conduct fusion experiments to verify this point. Specifically, we fusion our proposed CSST with the classical PQ~\cite{jegou2010product} and most recent proposed L\&C~\cite{douze2018link}. In addition, as the inspirational method catalyst~\cite{sablayrolles2019spreading} is close to our method in training networks to compress feature dimension, we compare our proposed model with catalyst on Deep1M, the dataset which is used in both this paper and their paper. Experiment results are listed in Table~\ref{Table:fusion}.

Experiment results show that using compressed feature learned by our proposed CSST improves both search accuracy and speed for PQ related methods. On Deep1M, in experiments of quantizing feature vectors to 32 bytes, combining CSST with PQ improves recalls 1@1, 1@5 and 1@50 by 21.0, 21.0 and 4.1 percentage points respectively. 
Our proposed CSST also brings more improvement than Catalyst. In experiments of coding input vectors with 16 bytes, using Catalyst improves recalls 1@1 and 1@5 by 3.0 and 9.3 percentage points and our proposed CSST improves recalls by 11.7 and 27 percentage points. 

Using CSST also improves search accuracy for L\&C, but the improvements CSST brings in here are less than that in PQ. This may result from that 2-level residual codec already optimized PQ quantized codes once and there is less room for improvement. Overall, CSST still improves search accuracy for L\&C, while keeping high search speed. 

\subsection{Working together with scalar quantization method}
In quantization methods, besides PQ related methods, scalar quantization methods are also widely used. In this section, we compare and fuse our proposed CSST with scalar quantization method proposed in~\cite{graydeploying}. 

From experimental results collected in Table~\ref{Table:scalar_quantization} we can see: 1) both scalar quantization and CSST can speed up indexing, but using CSST achieves faster indexing speed. This is because the distance of two 8-bit quantized codes still has to be represented by 16-bits, which reduces the data parallelism when distance of two quantized vectors are computed using AVX2 or AVX512 instructions. For example, computing the distance of two 16 dimensional float32 vectors (64 bytes) needs two AVX2 instructions calls, but computing the distance of two 64 dimensional 8-bit quantized vectors (also 64 bytes) needs to call the instruction 4 times.  2) Jointly using scalar quantization and CSST reduce indexing time the most. In short, scalar quantization method and our proposed CSST are complementary. 

\begin{table}[t]
\footnotesize
\renewcommand\arraystretch{1.2}
\setlength\tabcolsep{2.0pt}
\begin{center}
\caption{Experiments of working together with scalar quantization method. The compression factor is set to 4 and HNSW is used as baseline.}
\begin{tabular}{c|cc|c|ccc}
\hline
\multirow{2}*{Datasets} & \multirow{2}*{CSST} & Scalar  &  Indexing &  \multicolumn{3}{c}{Recall} \\
         &      &  Quant.  &  time (s) &  1@1  & 1@10  & 100 @ 100   \\
\hline
\multirow{4}*{GIST1M}
         & N  &  N & 733 & 97.40 & 100.00 & 94.39 \\
         & N  &  Y & 503 & 97.30 & 100.00 & 94.37 \\
         & Y  &  N & 245 & \bf{98.00} & 100.00 & \bf{95.27} \\
         & Y  &  Y & \bf{121} & 97.00 & 100.00 & 91.42 \\
\hline
\multirow{4}*{Deep1M}
         & N  &  N & 300 & 99.50 & 100.00 & 95.23  \\
         & N  &  Y & 148 & 99.70 & 100.00 & 95.19  \\
         & Y  &  N & 100 & \bf{99.80} & 100.00 & \bf{95.60}  \\
         & Y  &  Y & \bf{74}  & \bf{99.80} & 100.00 & 95.55  \\
\hline
\end{tabular}
\end{center}
\label{Table:scalar_quantization}
\end{table}

\begin{table}[h]
\footnotesize
\renewcommand\arraystretch{1.2}
\setlength\tabcolsep{2pt}
\begin{center}
\caption{Comparison with other dimension reduction methods. The compression factor is set to 4. Brute force denotes exhaustive search.}
\begin{tabular}{c|c|ccc|ccc}
\hline
    &        &  \multicolumn{3}{c|}{HNSW}    &  \multicolumn{3}{c}{Brute force} \\
\cline{3-8}
Datasets    & Methods &  \multicolumn{3}{c|}{Recall}  & \multicolumn{3}{c}{Recall}  \\  
            &         &  1@1  & 1@5  &  1@10  & 1@1 & 1@5 & 1@10 \\
\hline
\multirow{4}*{GIST1M}    
&   SRP &  24.9 & 49.4  &  60.7  &  24.8 & 49.7 & 60.9 \\
&   MLP &  47.9 & 73.8  &  84.2  &  47.9 & 73.8 & 84.1 \\
&   VAE &  49.2 & 77.0  &  86.0  &  49.3 & 76.8 & 85.8 \\
&   CSST &  \bf{80.1} & \bf{98.4}  &  \bf{99.8}  &  \bf{80.9} & \bf{98.3} & \bf{99.7} \\
\hline
\multirow{4}*{Deep1M} 
& SRP         &  17.3 & 35.6  &  44.4  & 17.3  & 35.6  & 44.4 \\
& MLP         &  48.3 & 77.6  &  89.1  & 48.4  & 77.6  & 88.9  \\
& VAE         &  50.0 & 81.9  &  90.2  & 50.0  & 81.9  & 90.2  \\
& Catalyst    &  57.9 & 89.6  &  92.1  & 57.9 & 89.6  &  92.1  \\
& CSST         &  \bf{67.3} & \bf{94.9}  &  \bf{98.9}  & \bf{67.3}  & \bf{94.9}  & \bf{98.9} \\
\hline
\end{tabular}
\end{center}
\label{Table:comparison}
\end{table}

\subsection{Comparison with other compression methods}
Previous experiments have shown that using compression feature vectors learned by our proposed CSST speeds up indexing without sacrificing accuracy. Besides our CSST, there are other compression methods. In this section, we compare the propsoed models with other methods to evaluate its efficiency. 

Here, we employed five comparison methods, including one traditional methods and four network based learning methods. Table~\ref{Table:comparison} presents comparison results. 

Compared with other four compression methods, our proposed CSST achieves the highest accuracy. For GIST1M, using CSST and HNSW achieves 80.1\% recall 1@1, 30.9 percentage points higher that that of VAE. As is analysed in the section \textit{the proposed method}, using single sparse random projection harms search accuracy seriously. Both VAE is classical and powerful compression method, but it focus on keeping information of the input feature instead of keeping the neighborhood structure, which is not aligned with the requirement of ANNS. They are outperformed by our proposed CSST. On Deep1M, the proposed CSST also achieves better performance, even compared with the most recent proposed method Catalyst.

\section{Discussion}

In this work, we have proposed a generic feature compression method for ANNS problem. The proposed method consists of a compression network  (CSST) which combines traditional projection function and transformer model, and an inhomogeneous neighborhood relationship preserving (INRP) loss which is aligned with the characteristic of ANNS. The proposed method can be generalized to most ANNS methods. It speeds up indexing speed to 2$\times$ to 4 $\times$ its original speed without hurting accuracy for graph based methods, such as NSG and HNSW. It improves recalls by several or even a dozen percentage points for PQ related methods. 

While demonstrated our model's overall effectiveness and potential usage, we have not yet verified the importance of individual design choices or the hypothesis behind our approach. We will leave these to our future work.

\bibliography{tech_report}
\bibliographystyle{tech_report}

\end{document}